\documentclass[aps,preprint,a4]{revtex4-1}

\usepackage{amssymb}
\usepackage{hyperref}
\usepackage{amsmath}
\usepackage{bm}
\usepackage{color}
\usepackage{epsfig}
\usepackage{float}
\usepackage{graphicx}

\begin{document}
	
\title{Is there a finite mobility for the one vibrational 
mode Holstein model? Implications from real time simulations} 

\author{Tianchu Li}
\author{Yaming Yan}
\author{Qiang Shi}\email{qshi@iccas.ac.cn}

\affiliation{Beijing National Laboratory for
Molecular Sciences, State Key Laboratory for Structural Chemistry of
Unstable and Stable Species, CAS Research/Education Center for
Excellence in Molecular Sciences, Institute of Chemistry, Chinese
Academy of Sciences, Zhongguancun, Beijing 100190, China}

\affiliation{University of Chinese Academy of Sciences,
Beijing 100049, China}

\begin{abstract}

The question of whether there exists a finite mobility
in the standard Holstein model with one vibrational mode on 
each site remains unclear. In this letter, we approach this 
problem by employing the hierarchical equation of motion (HEOM) 
method to simulate model systems where the vibrational modes are 
dissipative. It is found that, as the friction becomes smaller, 
the charge carrier mobility increases significantly and a 
friction free limit can not be obtained.
The current autocorrelation functions are also calculated for the 
friction free Holstein model, and converged results can not 
be obtained with the increase of the number of sites.
Based on these observations, we conclude that a finite mobility 
can not be defined for the standard Holstein model, 
which is consistent with the recent 
finding by Kolss {\it et al.} (Phys. Rev. Lett. 123, 126601).

\end{abstract}

\maketitle

%\section{Introduction}
Charge carrier mobility is an important property of organic 
semiconducting materials, and is critical in many of their
applications \cite{tsutsui16,gershenson06,coropceanu07}. 
It is now well known that 
charge carrier transport in these materials are 
significantly affected by the electron-phonon 
interaction \cite{holstein59,silbey80,silinsh94,troisi11}. 
Depending on the strengths of the intermolecular electronic 
coupling and the electron-phonon coupling,
the charge carrier transport mechanism 
can range from band-like behavior \cite{farchioni01, 
gershenson06} to hopping transitions \cite{nan09},
and cases in between \cite{troisi06,fratini17,li21}.
As a unified theory is not yet available, 
charge carrier transport mechanism in organic materials 
continues to be an active area of theoretical research.

The Holstein model\cite{holstein59,holstein59a} has been 
widely applied to study charge carrier transport 
in organic molecular crystals (OMCs), and has been 
indispensable in understanding such processes \cite{cheng08a,
fetherolf20}. Analytically solutions of the Holstein model 
are only available 
in the strong and weak coupling limit \cite{mahan00,lang63,
alexandrov10}. In the case of arbitrary coupling, 
approximate \cite{berciu06,zoli00,cataudella99}
and numerically exact \cite{mishchenko08,goodvin11,bonvca99, 
jeckelmann98,barivsic04} methods can be employed. 
For example, quantum Monte Carlo (QMC) \cite{kornilovitch98, 
romero98,romero99} simulation can be used to obtain imaginary 
time properties, while the charge carrier transport mobility 
can be obtained from an inverse Laplace transformation 
of the imaginary time results. 
The Green function methods\cite{langreth64,mahan00,goodvin11,
prodanovic19,bonvca19,mitric22,mishchenko15} can 
also be used to numerically simulate the Holstein model
and obtain the charge carrier mobility. 

Charge carrier mobility can also be obtained by calculating the 
real time current autocorrelation functions by using methods such as 
the density matrix renormalization group (DMRG) \cite{jeckelmann98,
jansen20,li20a} approach. Since long time simulations are still very 
challenging, the correlation functions are usually truncated at a 
finite time. For this treatment to be valid, it is required that,
either the long time behavior of the autocorrelation function 
is not important, or can be extrapolated using simple approximations.

A particularly interesting problem in the literature is 
the one vibrational mode Holstein model, where each 
site couples to a local vibrational mode.
It is still not very clear whether there is a well defined 
diffusion constant for this problem 
when there is no static disorder or dissipation.
In the literature, charge carrier mobilities have been obtained from 
both imaginary time and real time methods \cite{troisi06,troisi07,
wang11a,yan19,fetherolf20}.
However, in a recent study of the friction free Holstein model 
by Kloss et al. \cite{kloss19}, it is found that the 
mean square displacement (MSD) does not show a linear dependence 
of time $t$ for a wide range of parameters.
This result indicates that a diffusion 
constant may not exist and thus the charge carrier transport mobility 
can not be defined for the standard one vibrational 
mode Holstein model.
 
In previous works, our group has employed the hierarchical 
equation of motion (HEOM) method to simulate the Holstein or
Holstein-Peierls models \cite{wang10,song15,yan19}.
The HEOM approach is a ``numerically exact" method to simulate  
quantum dynamics in condensed phases \cite{tanimura89,
tanimura20}, and is widely applied in different fields 
including excitation energy 
transfer (EET) dynamics and related spectroscopic 
phenomena \cite{ishizaki09b,yan21}, as well as nonequilibrium 
charge carrier transport dynamics \cite{jin08}.
In all our previous studies of the Holstein model \cite{wang10,
song15,yan19}, dissipative 
modes were employed for the vibrational degrees of freedom (DOFs),
and well defined diffusion constants and mobilities can be obtained.

In this work, we investigate whether a well defined 
mobility exists for the 1D one vibrational mode Holstein model,
by starting from a 1D Holstein model with dissipation and 
systematically decreasing the dissipation strength towards 
the friction free limit. More specifically, we apply the methods 
previously developed in Refs. \cite{li22,xing22} to a Holstein 
model with a Brownian oscillator (BO) spectral density, 
where the strength of dissipation is controlled by a 
single parameter. Charge carrier mobilities for the 
dissipative model are then calculated from both the MSD 
and the real time current autocorrelation functions, 
and extrapolated to the friction free limit.

We start from the total Hamiltonian for a 
standard 1D single mode Holstein model, 
which is given by \cite{holstein59}:
\begin{align}
\hat H_S = \hat H_e + \hat H_{vib} + \hat H_{e-vib}\;\; .
\end{align}
Here, the electronic Hamiltonian $\hat H_e$ is defined as: 
\begin{align}
\label{eq:hsys}
\hat H_e = -J\sum_n\left ( {\hat c}^\dagger_{n+1}  {\hat c}_{n} 
+ {\hat c}^\dagger_{n}  {\hat c}_{n+1} \right )  \;\; ,
\end{align}
where the $\hat c^\dagger_n$ and $\hat c_n$ are the creation and 
annihilation operators of the electron at site $n$, $J$ is 
the electronic coupling between nearest neighbors.

The phonon Hamiltonian $\hat H_{vib}$ is given by:
\begin{align}
\label{eq:bath}
\hat H_{vib} = \Omega\sum_n {\hat b}^\dagger_n {\hat b}_n \;\; ,
\end{align}
where the $\hat b^\dagger_n$ and ${\hat b}_n$ are the creation 
and annihilation operators of the vibrational mode at site $n$ with 
frequency $\Omega$.

The interaction term is defined as,
\begin{align}
\label{eq:sb}
\hat H_{e-vib} = C\sum_n \left ( \hat b^\dagger_n
 + \hat b_n\right )\hat c^\dagger_n \hat c_n \;\;,
\end{align}
where $C$ is the coupling constant between the electronic DOF and 
the vibrational mode. 

As stated above, we first simulate a model system 
where the vibrational modes are subjected to dissipation.
To this end, each vibrational mode is coupled to 
a set of harmonic oscillators, such that the total bath Hamiltonian 
is written as:
\begin{align}
\hat H_B = \sum_n\sum_m \left ( \frac{\hat p_{n,m}^2}{2M_{n,m}} 
+ \frac{1}{2}M_{n,m}\omega_{n,m}^2\hat x_{n,m}^2\right )\;\; ,
\end{align}
where $\hat p_{n,j}$, $\hat x_{n,j}$, $M_{n,j}$, and $\omega_{n,j}$ 
are the momentum, position, mass, and frequency of the $j$th 
oscillator of $n$th site.  

The interaction between the vibrational modes
and harmonic bath is given by:
\begin{equation}
\hat H_I = \sum_n \sum_m\left ( {\hat b}^\dagger_n 
+{\hat b}_n\right )\hat x_{n,m} \;\;.
\end{equation}
The total Hamiltonian of the dissipative Holstein model is then
$\hat H_{tot} = \hat H_S + \hat H_B + \hat H_I$.
The vibrational part of 
the total Hamiltonian $\hat H_{vib} + \hat H_B$  + $\hat H_I$ 
can be diagonalized into a new set of harmonic oscillator modes:
\begin{equation}
\hat H'_{vib} = \sum_n\sum_j \omega'_{n,j}
\hat b'^{\dagger}_{n,j}\hat b'_{n,j} \;\;.
\end{equation}
By using this new sets of harmonic oscillator modes, 
the electron-phonon interaction term can be written as
\begin{equation}
\hat H^{\prime}_{e-vib} = \sum_n\hat F_n\hat c^\dagger_n\hat c_n\;\;,
\end{equation} 
where 
\begin{align}
\label{eq:F}
\hat F_n = \sum_j g_{n,j}( \hat b'^\dagger_{n,j}
+\hat b'_{n,j}) \;\;.
\end{align}
The spectral density for the $n$th site is defined 
as \cite{weiss12,garg85},
\begin{align}
\label{eq:spectraldensity}
J_n(\omega)=\frac{\pi}{2}\sum_ j\frac{g_{n,j}^2}{\omega'_{n,j}} 
\delta(\omega-\omega'_{n,j}) \;\;.
\end{align}

The above procedure to turn a problem from a system-vibrational 
mode-harmonic bath model into a new effective 
system-harmonic bath model has been introduced by 
Leggett, Garg $et \; al$. \cite{leggett84,garg85}.
In this work, we assume that the spectral density is the 
same for all sites, and takes the following form:
\begin{align}
\label{eq:BO}
J(\omega)=\lambda\frac{\Omega^2\gamma\omega}{(\omega^2-\Omega^2)^2 
+ 4\gamma^2\omega^2} \;\;.
\end{align}
The so called Brownian oscillator (BO) spectral density in 
Eq. (\ref{eq:BO}) can be derived by coupling a harmonic 
oscillator to an Ohmic bath \cite{leggett84,garg85,ito16,tanaka09}.

In Eq. (\ref{eq:BO}), the reorganization energy is determined by 
the vibrational frequency $\Omega$ and the coupling constant 
$C$ in Eqs. (\ref{eq:bath}) and (\ref{eq:sb}) as $\lambda = C^2/\Omega^2$.
So, the parameter $\gamma$ in Eq.(\ref{eq:BO}) actually 
controls the strength of the dissipative effects.
To study effects of dissipation on the charge carrier mobility,
we choose to fix the vibrational frequency $\Omega$ and 
reorganization energy $\lambda$, and investigate how 
$\gamma$ affects the charge carrier transport.
FIG. S1 in the supporting information shows 
the BO spectral density $J(\omega)$ 
for different values of $\gamma$, with $\lambda = 1.0$ and
$\Omega = 2.0$. 
It can be seen that the shape of $J(\omega)$ changes significantly
as $\gamma$ varies. When $\gamma = 0$, $J(\omega)$ is a delta 
function, corresponding to the standard friction free 
1D Holstein model. 
As $\gamma$ increases, $J(\omega)$ is broadened 
and the position of the peak shifts to lower frequencies. 
When $0 < \gamma <\Omega$, the system is underdamped,
and when $\gamma > \Omega$ the system is overdamped.

The charge carrier mobility of the dissipative Holstein model 
is obtained by using two different approaches. 
The first one is based on starting from an initial state and 
obtain the MSD using real time 
propagation: 
\begin{align}
{\rm MSD}(t)= \sum_{j=1}^N j^2P_{j}(t) \;\;,
\end{align}
where $P_j$ is the population of the $j$th site. 

The diffusion constant $D$ is then obtained when 
the MSD reaches the linearly growth region:
\begin{align}
D = \frac{1}{2}\lim_{t\to\infty}\frac{d{\rm MSD}(t)}{dt}\;\;.
\end{align}
The charge carrier mobility is then obtained 
by the Einstein relation: $\mu = eD/k_BT$ \cite{kubo95},
where $k_B$ is the Boltzmann constant, $T$ is the temperature.

Another approach to calculate the charge carrier mobility is
based on the Green-Kubo relation \cite{mahan00}:
\begin{align}
\label{eq:gk}
\mu = \frac{1}{2k_BT}\lim_{\omega\to 0}
\int_{-\infty}^{\infty}dte^{i\omega t}\langle \hat j(t)\hat j(0)\rangle\;\;,
\end{align}
where the current operator $\hat j$ in the Holstein 
model is given by
\begin{align}
\label{eq:jcurrent}
\hat j = -iJ\sum_m({\hat c}^\dagger_{m+1}  {\hat c}_{m} 
- {\hat c}^\dagger_{m}  {\hat c}_{m+1} )\;\;.
\end{align}

To obtain the current autocorrelation function 
$\langle \hat j(t)\hat j(0)\rangle$,  
we first use the imaginary time HEOM to 
calculate the correlated initial state, i.e., the equilibrium 
$\tilde\rho_{\bf n}$ . The current operator $\hat j$ is 
then multiplied to the equilibrium $\tilde\rho_{\bf n}$
and the new quantities $\hat j \tilde\rho_{\bf n}$
is used as the initial state to propagate the real time HEOM.
Finally, we multiple the current operator to the reduced density
operator at 
time $t$, and trace over the electron DOFs to obtain the 
correlation function $\langle \hat j(t)\hat j(0)\rangle$.  
The exact form and details of the real and imaginary time 
HEOM can be found in the supporting information.
 
By applying proper approximations, the current autocorrelation 
function can also be calculated analytically. When the 
electronic coupling constant $J$ is much smaller than the 
reorganization energy $\lambda$, 
by applying the second order perturbation with respect to $J$,
the current autocorrelation function can be calculated as \cite{wang11d}:
\begin{align}
\label{eq:fgrcc}
C(t) \approx 2J^2{\rm Re}\exp\{-g(t)\}\;\;,
\end{align}
where
\begin{align}
g(t)= \frac{2}{\pi\hbar}\int d\omega \frac{J(\omega)}{\omega^2}\{ 
\coth(\beta\hbar\omega/2)[1-\cos(\omega t)]+i\sin(\omega t) \} \;\;.
\end{align}
This equation is used to help analyzing the simulation 
results when $J$ is small.

By applying the real time HEOM, 
we first calculate the MSD for the dissipative Holstein model.
The periodic boundary condition is employed, 
and two different sets of parameter are used in the simulation.
In the first set of parameters, $\lambda = 1.0, \Omega=1.0, 
\beta = 1.0$, and $J=0.1$. Since the reorganization 
energy $\lambda$ is significantly larger than the electronic 
coupling constant $J$, it is supposed to be in the 
strong electron-phonon coupling regime. 
Five different values for the friction parameter $\gamma=$ 
$0.05, 0.2, 0.5, 0.8, 1.5$ are used in the simulation, 
where the first four are in the underdamped regime ($\gamma<1.0$), 
and the last one is in the the overdamped regime ($\gamma>1.0$). 

\begin{figure}[htbp]
\centering
\includegraphics[width=15cm]{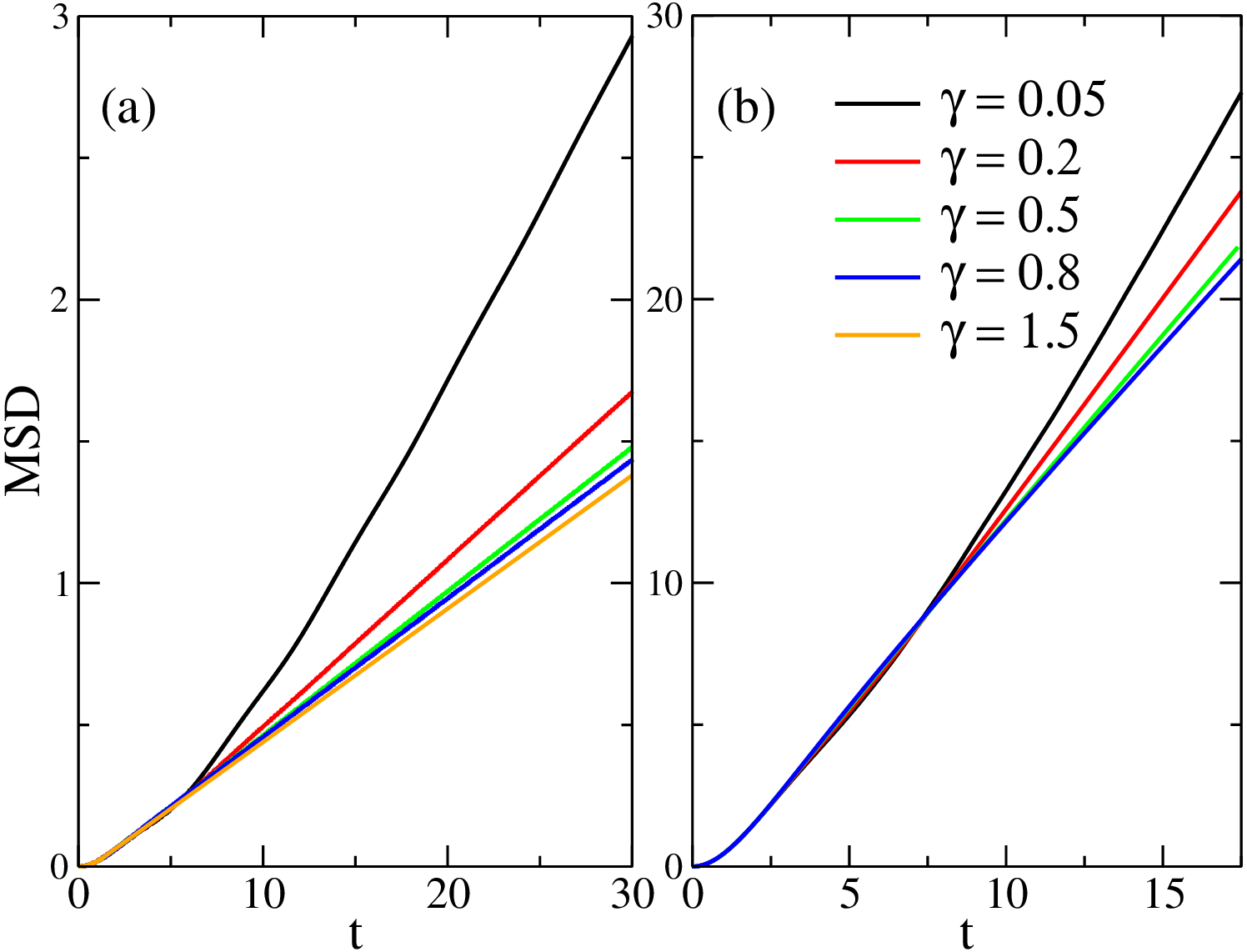}
\vspace{1em}
\caption{MSD simulated by the HEOM method for different
values of $\gamma$: (a) $J=0.1$, (b) $J =0.5$.
The other parameters are $\lambda$ = 1.0, $\Omega$ = 1.0,
and $\beta$ = 1.0.}
\label{fig1}
\end{figure}

The initial state of the real time HEOM simulation is 
$|0\rangle\langle0|\otimes e^{-\beta \hat H_{vib}}$. 
The total number of sites used in the simulation is 
up to 21 depending on different values of $\gamma$, ensuring that 
the MSD has reached the linear growth region and the 
calculated diffusion constant is not affected by boundary 
effects. The HEOM simulation is based on the matrix product state(MPS) method with 
bond dimension of 80. 
The simulated MSD is shown in Fig. \ref{fig1}(a). It can be seen 
that, with the increase of the the friction constant $\gamma$, 
the slope of the MSD curve becomes smaller. Besides, the time 
to reach the linear growth region of the MSD also 
increases with the decrease of $\gamma$. 

In another set of parameters, we use $\lambda = 1.0$, $\Omega=1.0$, 
$\beta = 1.0$, and $J=0.5$. Since the reorganization energy
$\lambda$ and the electronic coupling constant $J$ are now
comparable in magnitude, this is a case in the so called 
intermediate coupling regime.
As the coupling constant $J$ is now much larger than 
that in the first set of parameters, the real time HEOM 
propagation becomes more challenging. To obtain converged results, 
the number of sites needed is 51 for $\gamma=1.5$, and
71 for $\gamma=0.05$. Bond dimension for the MPS also 
increases to 120.
The time dependent MSD obtained for the second set of parameters
is presented in Fig. \ref{fig1}(b). Same as the $J=0.1$ case, 
for smaller $\gamma$ values, it takes a longer time for the MSD to 
reach the linear growth region. The diffusion 
constants are also larger for small $\gamma$. 

We then apply the imaginary and real time HEOM to simulate the 
current autocorrelation function for the aforementioned two 
sets of parameters. The MPS method is applied to propagate 
imaginary time HEOM, with bond dimension of 90 for all 
different $\gamma$s. 
The cosine fitting scheme with seven cosine terms is 
used to fit the imaginary time bath correlation function \cite{xing22}. 
To obtain converged results, the numbers of sites used in the 
simulations are from 8 ($\gamma =0.8$) to 10 ($\gamma =0.05$) for 
$J=0.1$, and 10 ($\gamma =0.8$) to 26 ($\gamma =0.05$)  for $J=0.5$.
FIG. S3 in the supporting information shows the convergence of the 
current autocorrelation function with respect to different 
number of sites when $\gamma=0.05$, $J$=0.5. 

\begin{figure}[htbp]
\centering
\includegraphics[width=15cm]{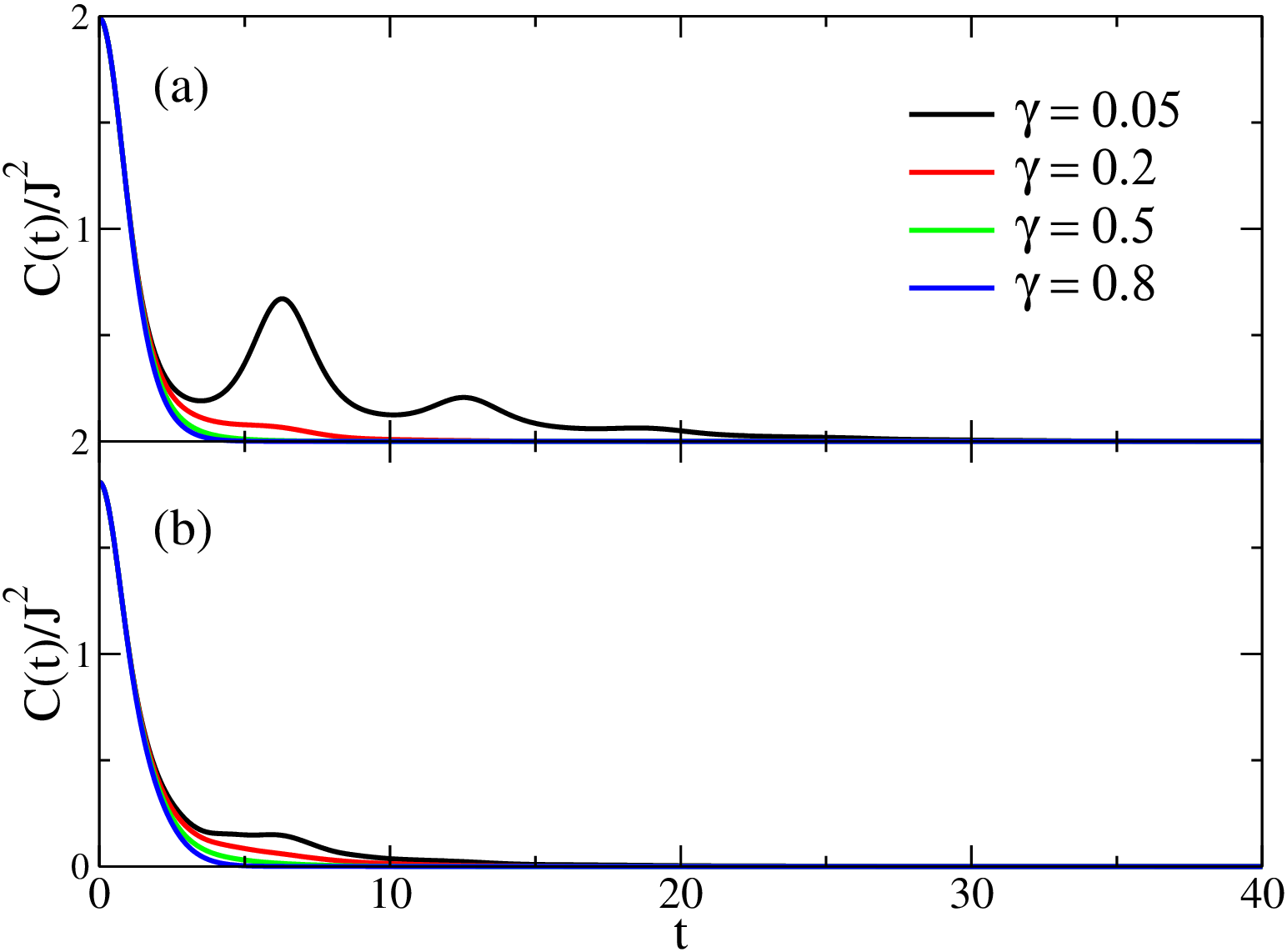}
\vspace{1em}
\caption{Current autocorrelation function
simulated by the HEOM method for different values of $\gamma$: 
(a) $J=0.1$, (b) $J =0.5$.
%The other parameters are $\lambda = 1.0, \Omega = 1.0,\beta = 1.0$.
}
\label{fig2}
\end{figure}

In the $J=0.1$ case, the current autocorrelation functions 
for different values of $\gamma$ are shown in Fig. \ref{fig2}(a), 
and the comparison with the second order perturbation is shown 
in Fig. S4 in the supporting information. The current 
autocorrelation functions for $J$ = 0.5 
are shown in Fig. \ref{fig2}(b) and Fig. S5. In both cases,
faster decay of the correlation function is observed when 
increasing $\gamma$. It is also shown that, for larger coupling 
constant $J$, the current autocorrelation function becomes less 
oscillatory. 

For all values of $\gamma$ and $J$, the current autocorrelation 
function decays to zero, which indicates that a 
finite mobility can be obtained via the Green-Kubo relation 
in Eq. (\ref{eq:gk}). The second order perturbation results in 
Fig. S4 and Fig. S5 also show that, 
although the accuracy of the second order autocorrelation 
functions depends on specific model parameters, 
they also decay to zero and leads a finite approximate mobility.

\begin{figure}[htbp]
\centering
\includegraphics[width=15cm]{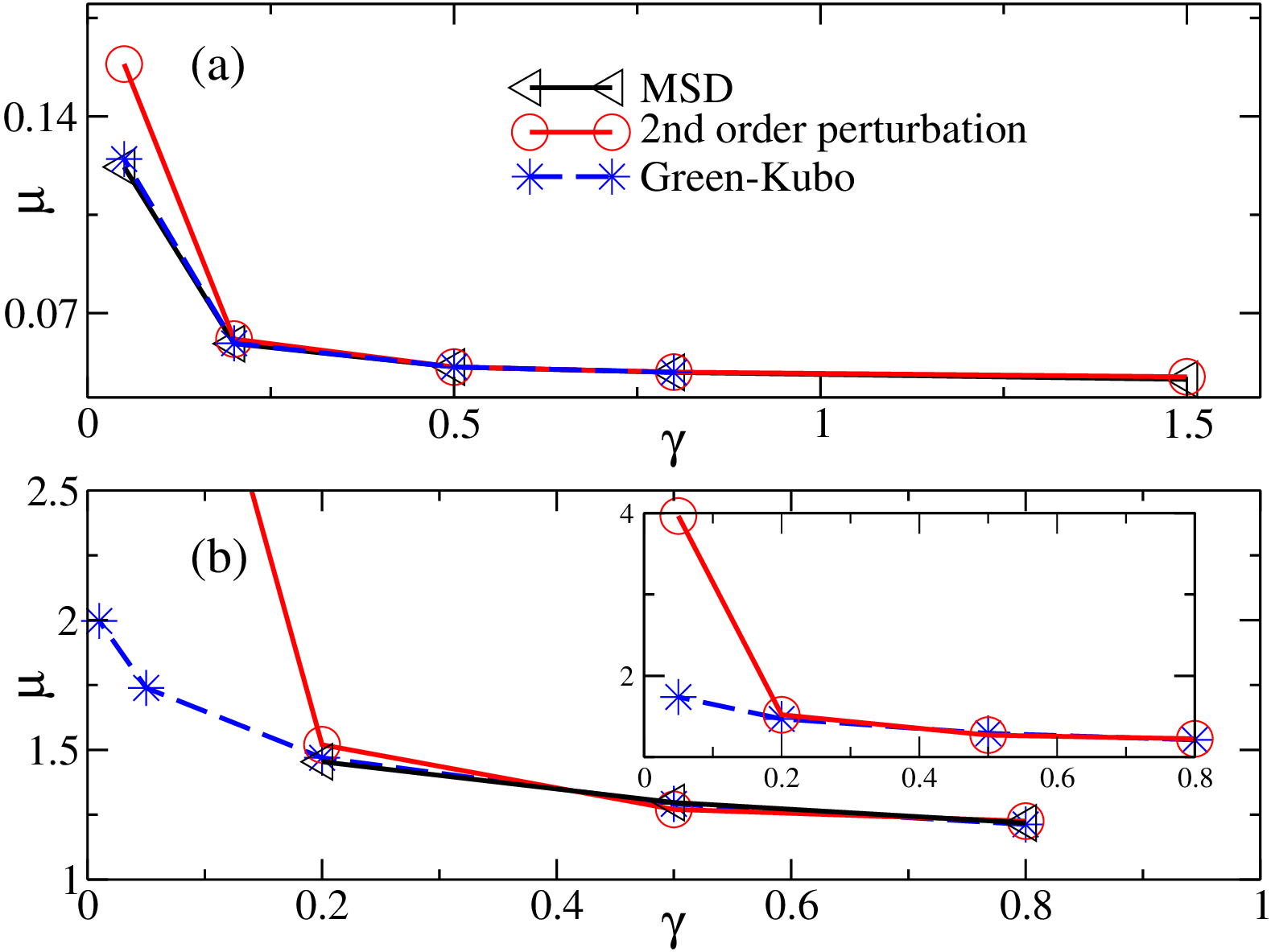}
\vspace{1em}
\caption{The mobility simulated by the HEOM method using the MSD
and current autocorrelation function, compared with the second
order perturbation result. (a) $J=0.1$, (b) $J =0.5$.
}
\label{fig3}
\end{figure}

The mobilities calculated by using the MSD method and 
the Green-Kubo relation are consistent with each other.
Fig. \ref{fig3}(a) shows the calculated mobilities
via the MSD and the Green-Kubo relation,
compared with the second order perturbation theory for $J=0.1$.
It can be seen that, the mobility decreases when the 
dissipation becomes stronger. The second order perturbation 
actually works very well for large frictions, but
overestimates the mobility at low friction. The same behavior 
is observed in the results for $J=0.5$ shown in Fig. \ref{fig3}(b),
although the overestimation is rather pronounced for small
friction ($\gamma = 0.05$).
It is thus shown that the friction constant $\gamma$ plays an 
important role in charge transport.  Increasing $\gamma$ 
leads to stronger decoherence, 
which makes the current autocorrelation  
function decaying faster to zero, and reduces the mobility. 

However, when we try to extrapolate the results to the friction free 
limit by taking $\gamma \rightarrow 0$, the $\gamma=0$
limit can not be defined in both the $J=0.1$ and 0.5 cases 
because of the sharp decrease of the mobility for small $\gamma$s.
I.e., the mobility tends to diverge when $\gamma\rightarrow 0$.
The divergence of the second order perturbation result is 
actually easy to understand: As shown in Fig. S6, 
in the case $\gamma =0$, the second order 
current autocorrelation function calculated via Eq. (\ref{eq:fgrcc})
a periodic function with the period of $2\pi/\Omega$, 
so the integration in the Green-Kubo relation goes to infinity.

It would then be interesting to investigate how the current
autocorrelation functions in the friction free Holstein model 
behave beyond the second order perturbation.
We now apply the combined imaginary and real time HEOM 
towards this problem. For the friction free Holstein model,
the imaginary time bath (vibrational) correlation function 
is expanded analytically by hyperbolic 
functions \cite{xing22}.

\begin{figure}[htbp]
\centering
\includegraphics[width=15cm]{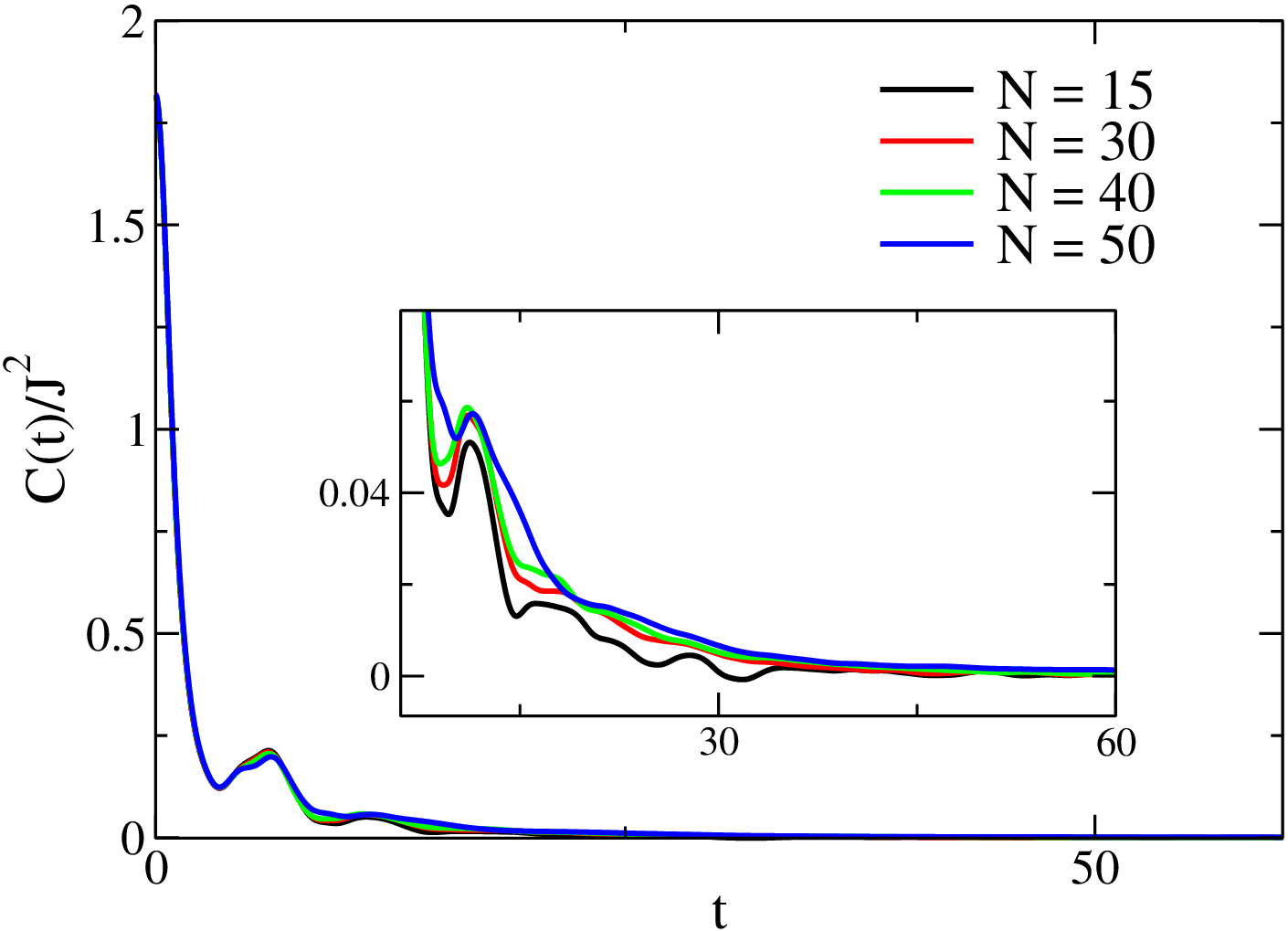}
\vspace{1em}
\caption{Current autocorrelation functions
simulated by the HEOM method for the friction free Holstein
model. The parameters used in the simulations are
$\lambda$ = 1.0, $\Omega$ = 1.0, $J$ = 0.5, and $\beta$ = 1.0.}
\label{fig4}
\end{figure}

The current autocorrelation functions for the friction free 
model with different number of sites are shown in Fig. \ref{fig4}. 
The parameters used in the simulation are 
$\lambda = 1.0$, $\Omega=1.0$, $\beta = 1.0$, $\gamma =0$, 
and $J =0.5$.
It can be seen that, the HEOM current autocorrelation function 
does not converge even for a large number of 
sites ($N$=50). 
However, compared with the second order result shown in Fig. S6,
the HEOM correlation function also does not show significant 
recurrences. So, based on above characteristic of the 
friction free current autocorrelation
function, and the failure to extrapolate the mobilities 
in the dissipative model 
to the zero friction limit, we conclude that a finite 
mobility may not be obtained for the friction free one vibrational mode
Holstein model.

In summary, we have applied the imaginary and 
real time HEOM to simulate the MSD and current 
autocorrelation function for the 1D single mode Holstein model. 
The main observation is that, with dissipative vibrational 
modes, the current autocorrelation function decays to zero, 
such that a finite mobility can be obtained. 
On the other hand, in the friction free
Holstein model, the current autocorrelation function 
does not decay to zero, and a finite mobility cannot be defined 
either from the Green-Kubo relation, or via extrapolation from 
finite friction results. This is consistent with the recent
finding that no diffusive behavior can be observed 
from the MSD of the friction free Holstein model \cite{kloss19}.
Of course, in realistic systems, dissipation and static 
disorder\cite{chuang21} are unavoidable, and a finite 
mobility should always be obtained.

\acknowledgments
This work is supported by NSFC (Grant No. 21933011).

\bibliography{../quantum}

\pagebreak

\pagebreak

\pagebreak

\pagebreak

\end{document}

\title{Supporting information for:
Is there a finite mobility for the one vibrational mode
Holstein model? Implications from real time simulations}
\author{Tianchu Li}
\author{Yaming Yan}
\author{Qiang Shi}\email{qshi@iccas.ac.cn}

\affiliation{Beijing National Laboratory for
Molecular Sciences, State Key Laboratory for Structural Chemistry of
Unstable and Stable Species, CAS Research/Education Center for
Excellence in Molecular Sciences, Institute of Chemistry, Chinese
Academy of Sciences, Zhongguancun, Beijing 100190, China}
\affiliation{University of Chinese Academy of Sciences,
Beijing 100049, China}

\maketitle

\section{The real and imaginary time HEOM for the dissipative Holstein model}
For the BO spectral density in Eq. (11),
\begin{align}
\label{eq:BO}
J(\omega)=\lambda\frac{\Omega^2\gamma\omega}{(\omega^2-\Omega^2)^2 
+ 4\gamma^2\omega^2} \;\;,
\end{align}
the bath correlation function can be decomposed into 
summation of exponential terms, 
\begin{equation}
\label{eq:ctsum}
C(t)=d_{+}\ e^{ -(\gamma+i\omega_{0})t}
+d_{-}\ e^{ -(\gamma-i\omega_{0})t}
+\sum_{k=1}^{\infty} d_k\ e^{-\nu_kt}\;,
\end{equation}
where $\nu_k=2\pi k/\hbar\beta$, and
\begin{subequations}
\begin{equation}
d_{\pm}=\frac{\hbar\lambda\Omega^2}{2\omega_{0}}
\left[\coth\frac{\beta\hbar(\omega_{0}\mp i\gamma)}{2}
\pm 1\right]\;,
\end{equation}
\begin{equation}
d_{k}=-\frac{8\lambda\Omega^2\gamma }{\beta}
\frac{\nu_k} {(\Omega^2+\nu_k^2)^2-4\gamma^2\nu_k^2} \;\;.
\end{equation}
\end{subequations}
the real-time HEOM for 
the dissipative Holstein model with the
BO spectral density is given by:
\begin{eqnarray}
\label{eq:heom}
\frac{\partial}{\partial t}\rho^{\mathbf{J}}_{\bf m,n}(t)
& = &-\frac{i}{\hbar}\left[ H_e,\rho^{\mathbf{J}}_{\bf m,n}\right]
-\left[\sum_{j = 1}^{N}(m_j+n_j)\gamma
+\sum_{j = 1}^{N}i(m_j-n_j)\sqrt{\Omega^2-\gamma^2}
+\sum_{j=1}^N\sum_{k=1}^{\infty}J_{j,k}\nu_k\right]
\rho^{\mathbf{J}}_{\bf m,n}\nonumber \\
& &-\frac{i}{\hbar}\sum_{j = 1}^{N}m_{j}\Theta_{j}^{+}
\rho^{\mathbf{J}}_{{\bf m}^{-}_{j},\bf{n}}
-\frac{i}{\hbar}\sum_{j = 1}^{N}n_{j}\Theta_{j}^-
\rho^{\mathbf{J}}_{{\bf m},{\bf n}_{j}^{-}}
-\frac{i}{\hbar}\sum_{j=1}^N\sum_{k=1}^{\infty}d_{j,k}J_{j,k}\left[|j\rangle\langle j|,
\rho^{\mathbf{J}_{j,k}^-}_{\bf m,n}\right]\nonumber \\
& &-\frac{i}{\hbar}\sum_{j = 1}^{N}
\left[ |j\rangle\langle j|,\rho^{\mathbf{J}}_{{\bf m}_{j}^{+},\bf n}
+\rho^{\mathbf{J}}_{{\bf m},{\bf n}_{j}^{+}}\right]
-\frac{i}{\hbar}\sum_{k=1}^{K}
\left[ |j\rangle\langle j|,\rho^{\mathbf{J}_{j,k}^+}_{\bf m,n}\right]\;,
\end{eqnarray}
where $\rho^{\bf 0}_{{\bf 0},{\bf 0}}$ 
with ${\bf m} = {\bf n} =0$, ${\bf J=0}$
is the reduced density operator (RDO), and
the other $\rho^{\mathbf{J}}_{\bf m,n}$ terms 
are auxiliary density operators (ADOs). The subscripts 
${\bf m}_{j},{\bf n}_{j}$ 
denote the primary BO mode on the $j$th site, while the 
superscript $\bf J$ denotes 
the Matsubara modes, ${\bf J}_{j,k}$ is for 
$k$th effective mode of the $j$th site. The detailed definition 
of the operators $\Theta^{\pm}$
can be found in previous works\cite{tanaka09,tanaka10,liu14,li22}. 
In this work, we also apply the Ishizaki-Tanimura closure 
method\cite{ishizaki05,shi09b} for the above HEOM by replacing the 
rapidly decaying factors $e^{-v_{k}t}$ by $\delta(t)/v_{k}$ 
for all Matsubara modes with $k>K$, where $K$ can be regarded 
as a control parameter for numerical convergence. The total number 
of the real time indices is thus $K_R =  K +2$ for each site. 
For convenience, we 
rewrite the RDO/ADOs $\rho_{\bf m,n}^{\bf J}$ into $\tilde\rho_{\bf n}$, 
where $\bf n$ stands for the real time indices including 
all the BO and Matsubara terms.

To obtain the equilibrium correlation functions of the Holstein model,
it necessary to obtain the correlated system-bath equilibrium 
state for the initial state of the real-time 
HEOM \cite{tanimura14,song15b,xing22,zhang22}.
In this work, we apply the imaginary time HEOM method 
in Refs.\cite{tanimura14,song15b,xing22} for this purpose. 
In this approach, 
the imaginary time correlation function is expanded into 
summation of cosine functions:
\begin{equation}
\label{eq:correxpand_cos}
\alpha_{\beta}(\tau) = \tilde c_0 + \sum_{k} 
\tilde c_{k}\cos\left[\tilde \gamma_k
\left(\frac{\beta}{2}-\tau \right)  \right]\;\;,
\end{equation}
and the imaginary time HEOM is given by 
\begin{align}
\label{eq:heqinit}
\frac{d\rho_{{\bf s},{\bf m},{\bf m'};{\boldsymbol{n}}}}{d\tau} 
&= -\hat H_e\rho_{{\bf s},\bf m,m';{\boldsymbol{n}}}-\sum_{j=1}^N\sum_{k}m_{j,k}\tilde\gamma_{k} 
\rho_{{\bf s},{\bf m}_{j,k}^{-},{\bf m'}_{j,k}^{+};{\boldsymbol{n}}}
+\sum_{j=1}^N\sum_{k}m'_{j,k}\tilde\gamma_{k} 
\rho_{{\bf s},{\bf m}_{j,k}^{+},{\bf m'}_{j,k}^{-};{\boldsymbol{n}}}\nl
&+\sum_{j=1}^N\sum_{k}{\tilde c}_k |j\rangle\langle j|
\rho_{{\bf s},{\bf m}_{j,k}^{+},{\bf m'};{\boldsymbol{n}}}
+\sum_{j=1}^N\sum_{k}\tilde{c}_k|j\rangle\langle j|
\rho_{{\bf s},{\bf m},{\bf m'}_{j,k}^{+};{\boldsymbol{n}}}\nl
&+ \sum_{j=1}^N\sum_{k} m_{j,k}|j\rangle\langle j|
\rho_{{\bf s},{\bf m}_{j,k}^{-},{\bf m'};{\boldsymbol{n}}}
+\sum_{j=1}^N\tilde c_0|j\rangle\langle j|\rho_{{\bf s}^+_j,{\bf m},{\bf m'};{\boldsymbol{n}}}
+ \sum_{j=1}^Ns_j|j\rangle\langle j|\rho_{{\bf s}_j^-,{\bf m},{\bf m'};{\boldsymbol{n}}} \nonumber \\
&-\sum_{j=1}^N\sum_{k=0}^{\infty}n_{j,k}|j\rangle\langle j|e_k(\tau)
\rho_{{\bf s},{\bf m},{\bf m'};{\boldsymbol{n}}_{j,k}^-}\;\; .
\end{align}
In the above Eq. (\ref{eq:heqinit}), 
$\rho_{s,\bf m,m';{\boldsymbol{n}}}$ is the imaginary time RDO/ADO, 
where $s,{\bf m}, {\bf m}'$ are the imaginary time indices, and $\bf n$
is the real time index. The number of real time indices in ${\bf n}$ is 
consistent with aforementioned $K_R$, and the number of imaginary 
time indices is defined as $K_\beta$.
The exact form of $e_k(\tau)$ can be found in Ref.\cite{xing22}.
To obtain the correlated initial state that is used in the 
subsequent real-time calculation, 
the imaginary-time HEOM in Eq. (\ref{eq:heqinit}) 
are first propagated from 0 to $\beta$. We then set the imaginary
time indices $s,{\bf m}, {\bf m}'=0$, and obtain the correlated 
equilibrium RDO/ADOs by dividing the $\rho_{0,{\bf 0},{\bf 0},{\bf n}}$ 
with the partition function $Z$, 
i.e., $\rho_{\bf n}^{eq} =\rho_{0,{\bf 0},{\bf 0},{\bf n}}/Z$.
where $Z = {\rm Tr}\rho_{0,{\bf 0},{\bf 0},{\bf 0}}$.

For the Holstein model, the dimension of the RDO/ADO indices $\bf n$ 
can become rather large when there are many sites. Although 
the on-the-fly filtering method\cite{shi09b}
can efficiently reduce the computation cost for the HEOM,
calculations for a system with many sites
become difficult. We then apply the tensor network method to 
propagate the HEOM. 
For the real time and imaginary time HEOM, the RDO/ADOs
$\tilde\rho_{\bf n}$ and $\rho_{s,{\bf m},{\bf m'};
{\boldsymbol{n}}}$ can be represented with the matrix 
product state (MPS) form:
\begin{equation}
\rho_{\mathbf{n}}^{i j} \approx \sum_{i_{1}, 
\ldots, i_{K_{tot}}} B_{0}\left(i, i_{1}\right) B_{1}
\left(i_{1}, n_{1}, i_{2}\right) \cdots B_{K_{tot}}\left(i_{K_{tot}-1}, 
n_{K_{tot}}, i_{K_{tot}}\right) B_{K_{tot}+1}
\left(i_{K_{tot}}, j\right) \;\;.
\end{equation}
Here, the total number of effective modes is $NK_R + 2$ 
for the real time HEOM, and $K_{tot} = N(K_\beta + K_R) +2$ 
for the imaginary time HEOM.

\bibliography{../quantum}

\pagebreak
\begin{figure}[htbp]
\centering
\includegraphics[width=15cm]{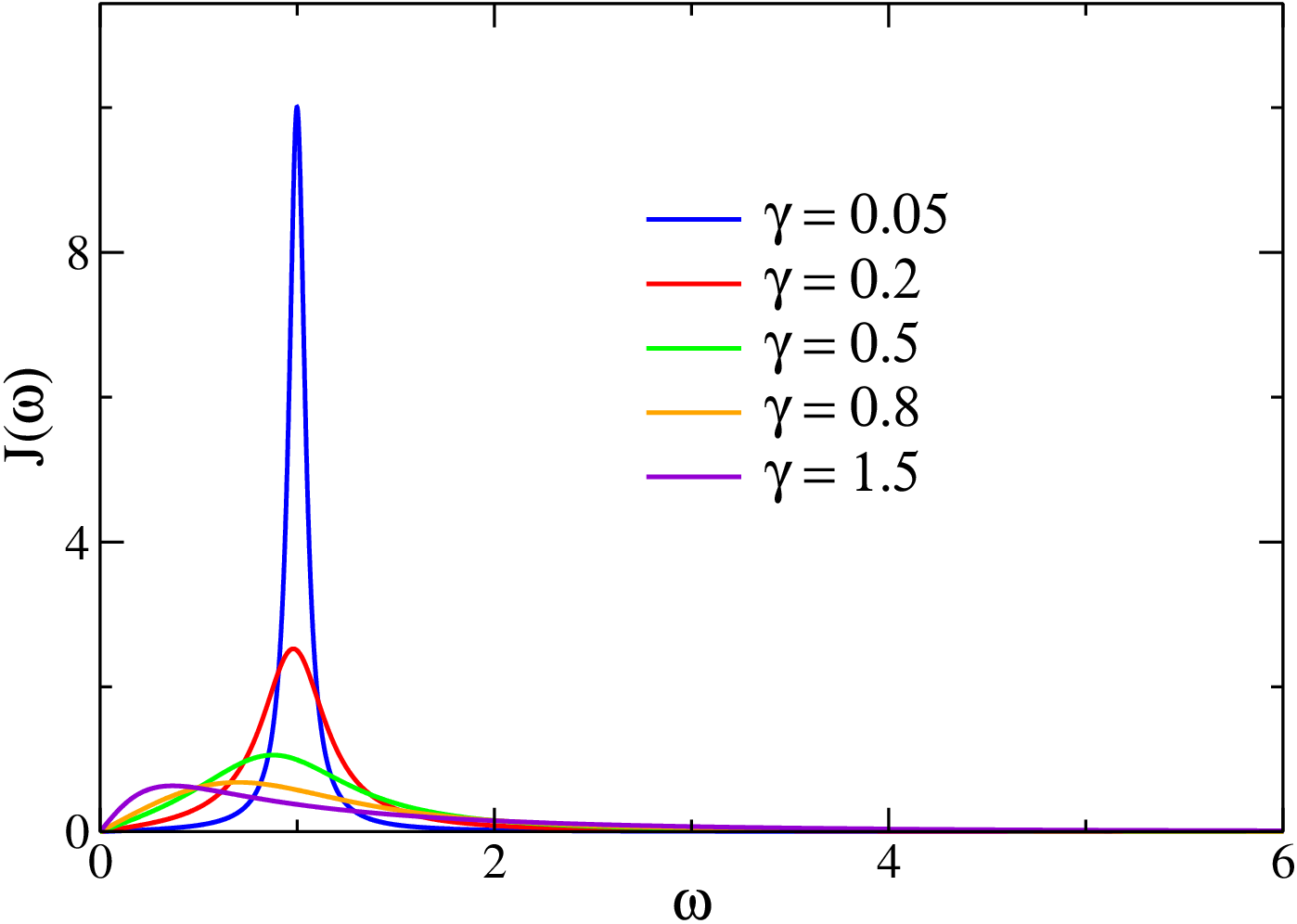}
\vspace{1em}
\caption{The shape of the BO spectral density in Eq. (\ref{eq:BO})
as a function of the friction parameter $\gamma$. In this figure, 
the reorganization energy $\lambda=1.0$ and the vibrational 
frequency $\Omega = 1.0$.}
\label{figs1}
\end{figure}

\pagebreak
\begin{figure}[htbp]
\centering
\includegraphics[width=15cm]{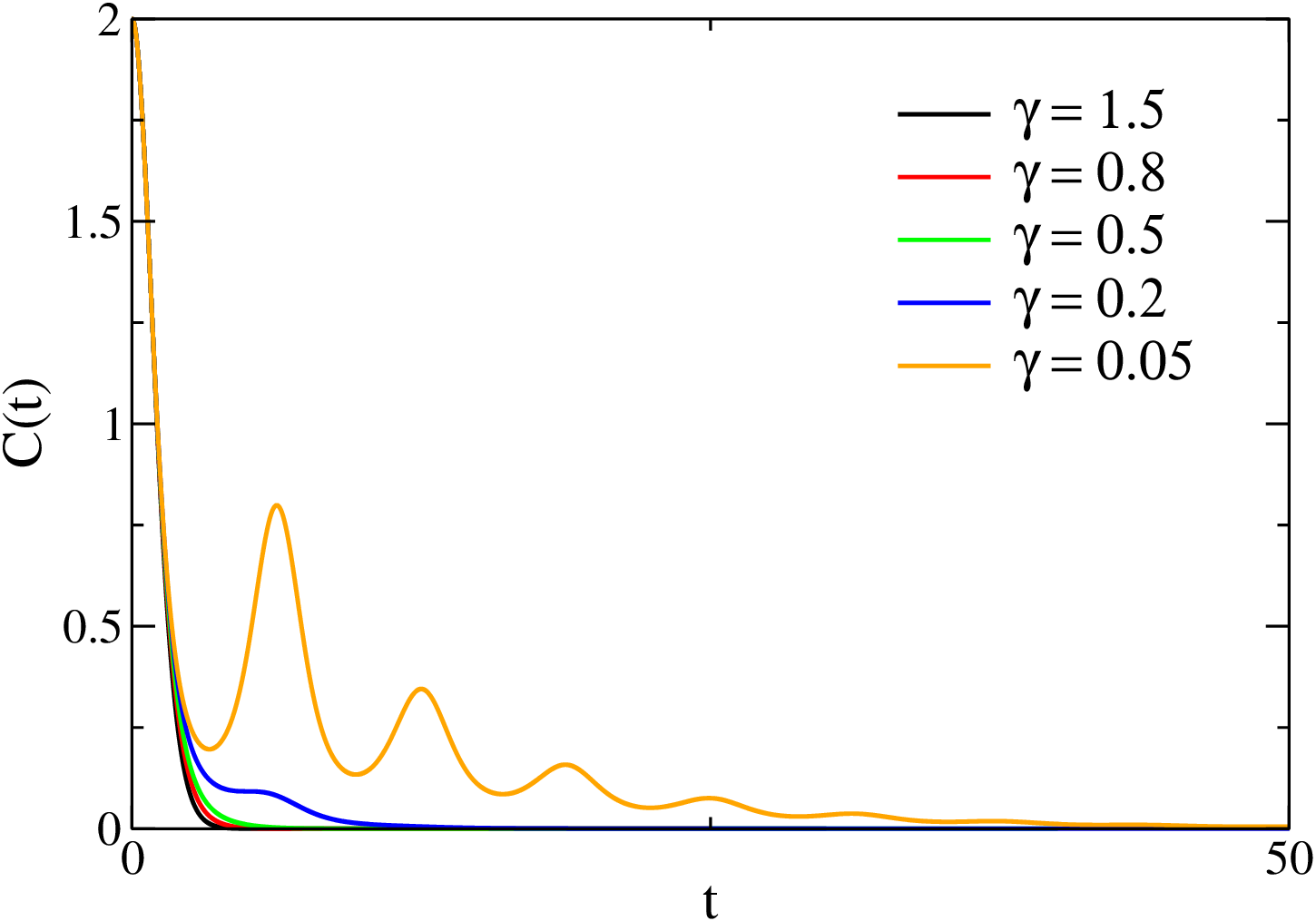}
\vspace{1em}
\caption{Current autocorrelation functions 
simulated from the second order perturbation. 
The parameter used in the simulations 
are $\lambda = 1.0, \Omega = 1.0, J = 0.1, \beta = 1.0$.}
\label{figs2}
\end{figure}

\pagebreak
\begin{figure}[htbp]
\centering
\includegraphics[width=15cm]{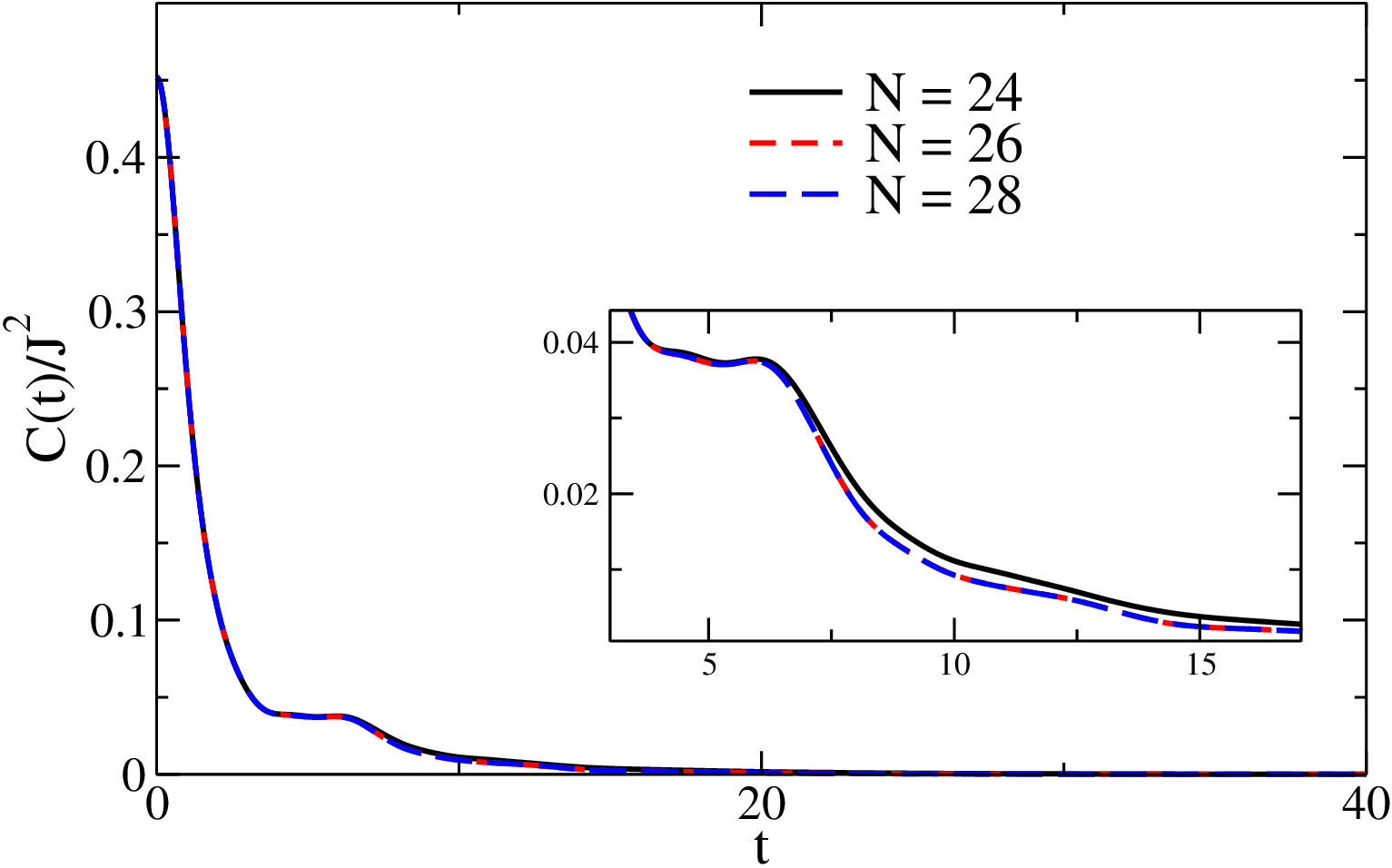}
\vspace{1em}
\caption{The current autocorrelation function 
simulated by the HEOM method for different number of sites 
with a small friction constant $\gamma = 0.05$. The parameter 
used in the simulations are $\lambda = 1.0, \Omega = 1.0, 
J = 0.5, \beta = 1.0$.
}
\label{figs3}
\end{figure}

\pagebreak
\begin{figure}[htbp]
\centering
\includegraphics[width=15cm]{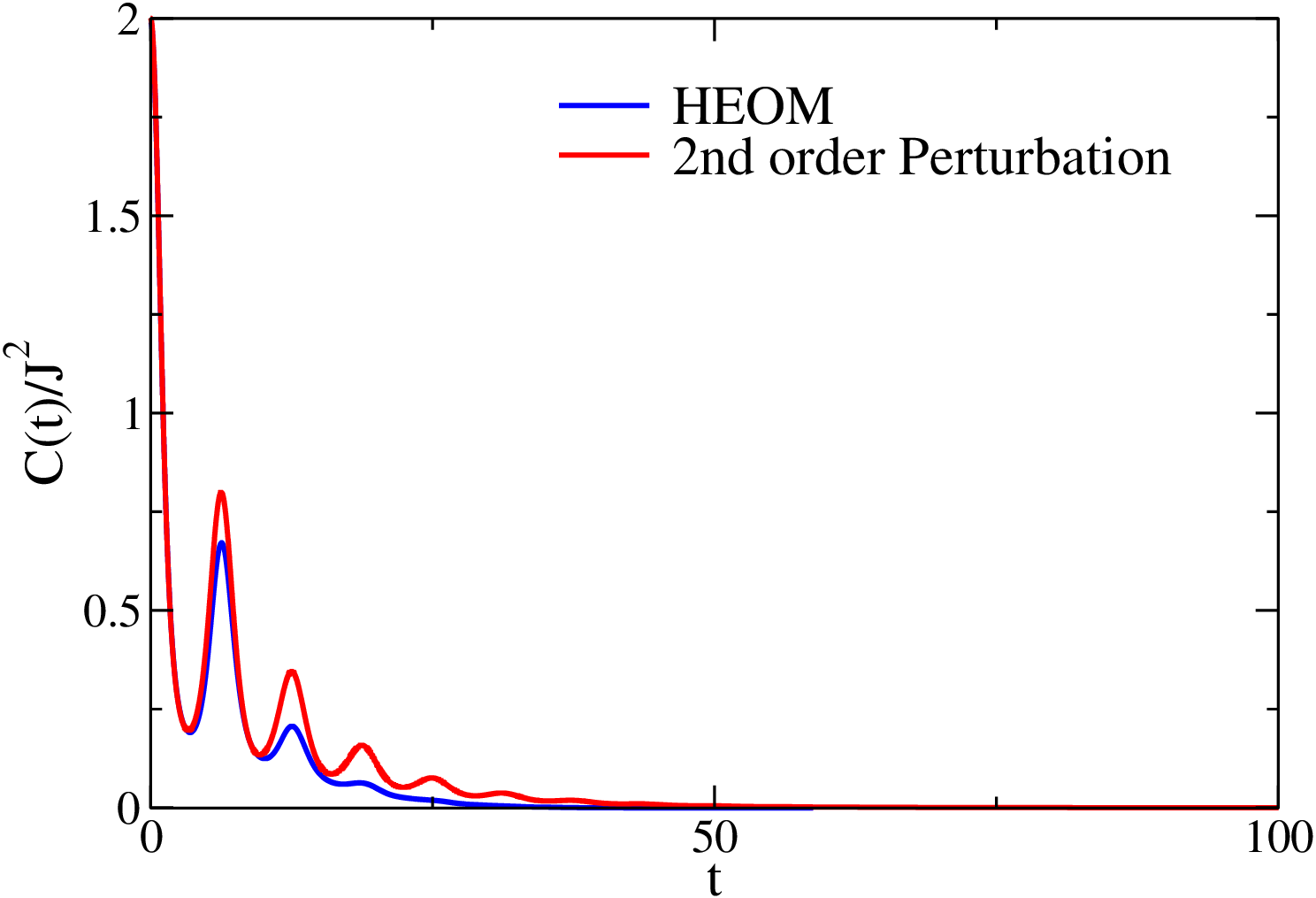}
\vspace{1em}
\caption{Comparison between the current autocorrelation function 
simulated by the HEOM method and the second order perturbation 
for a small friction constant $\gamma = 0.05$. The parameter 
used in the simulations are $\lambda = 1.0, \Omega = 1.0, 
J = 0.1, \beta = 1.0$.}
\label{figs4}
\end{figure}

\pagebreak
\begin{figure}[htbp]
\centering
\includegraphics[width=15cm]{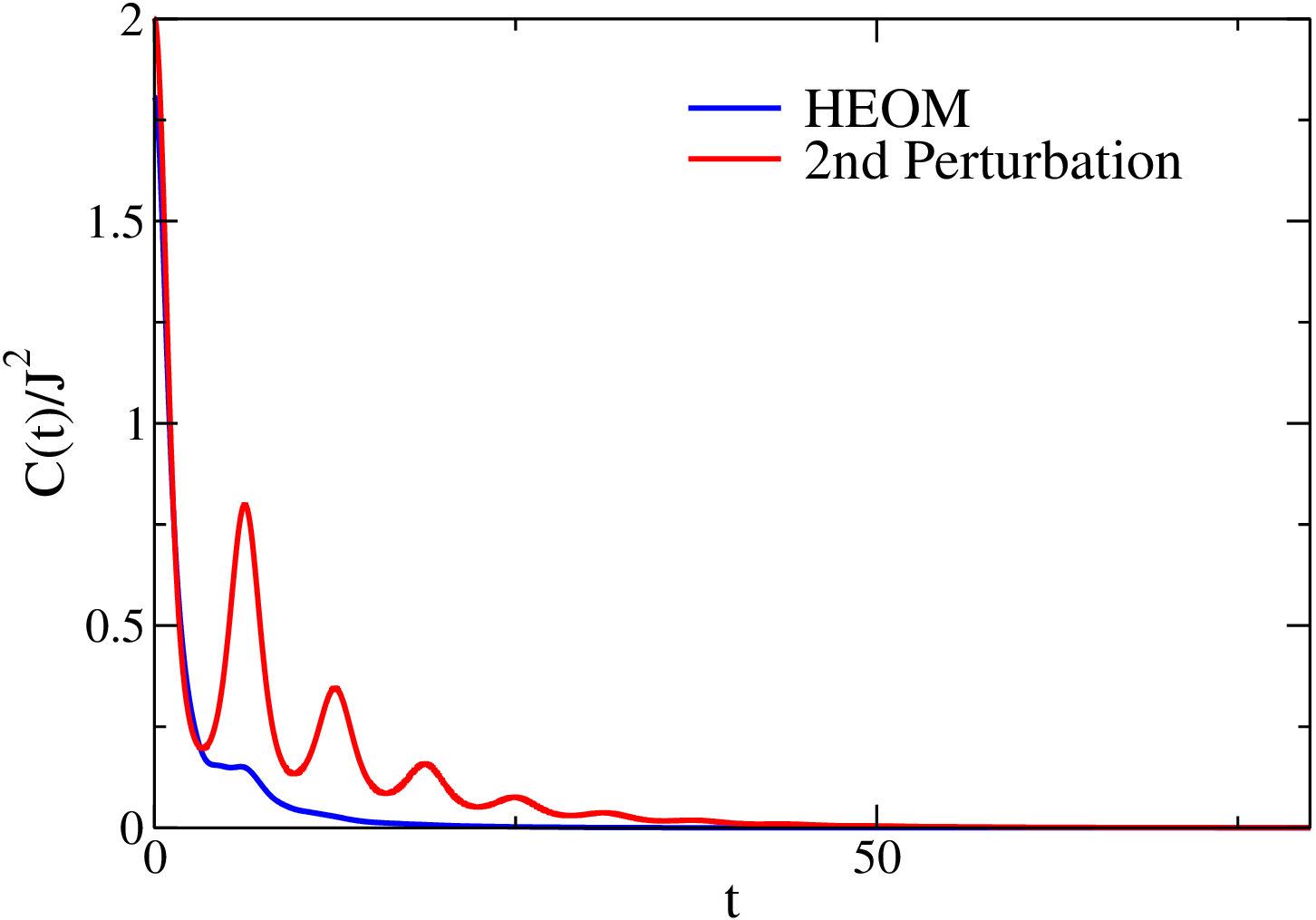}
\vspace{1em}
\caption{
Same as Fig. \ref{figs4}, for $J = 0.5$.
}
\label{figs5}
\end{figure}

\pagebreak
\begin{figure}[htbp]
\centering
\includegraphics[width=15cm]{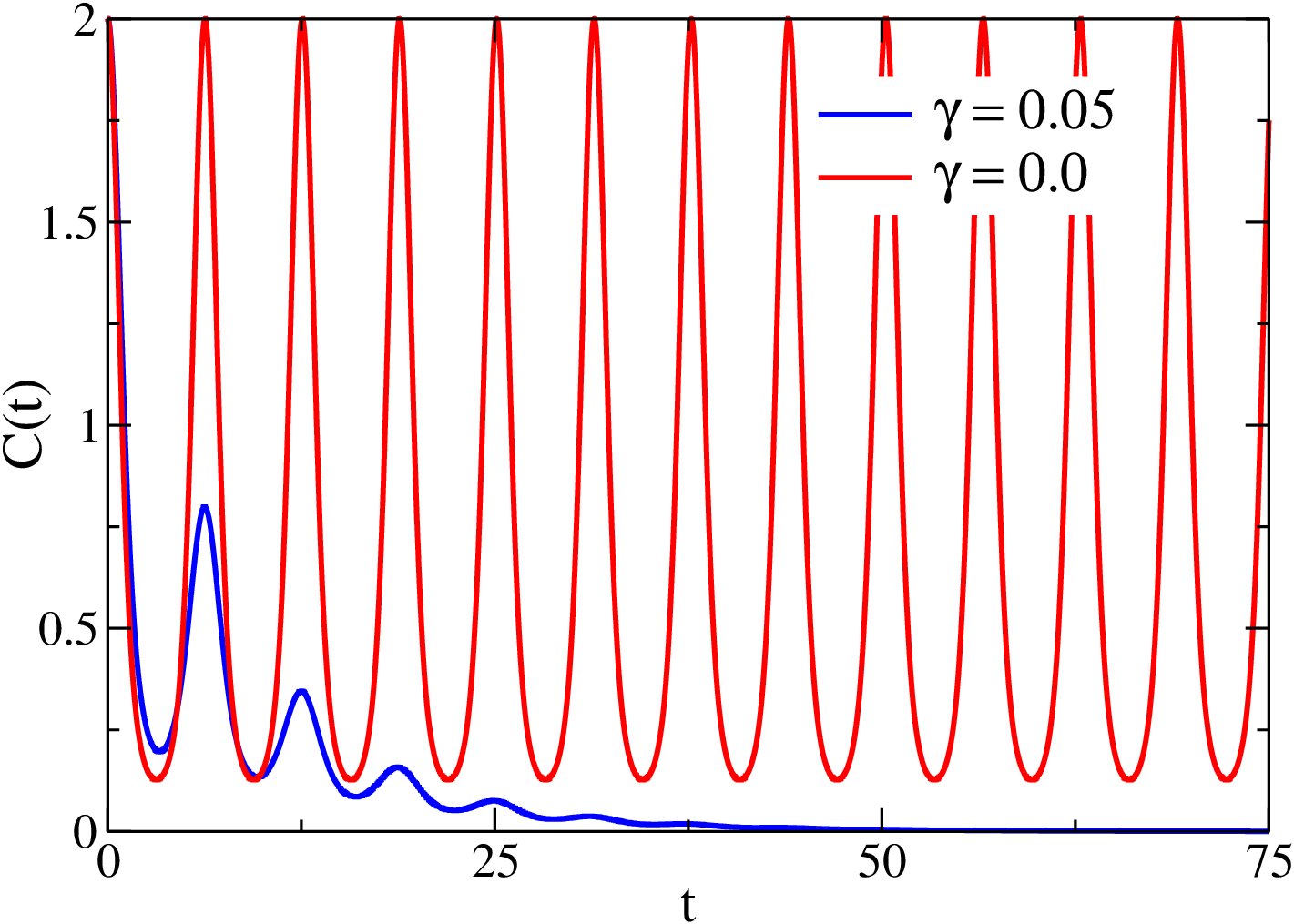}
\vspace{1em}
\caption{The second order current autocorrelation functions 
for different $\gamma$ and $\lambda$. The blue line is for the 
$\gamma = 0.8$ case, and the red line for the 
friction free case. The parameter used in the 
simulations are $\Omega = 1.0, J = 0.5, \beta = 1.0$, 
$\lambda = 1.0$.}
\label{figs6}
\end{figure}